# Timoshenko nonlocal strain gradient nanobeams: variational consistency, exact solutions and carbon nanotube Young moduli

Raffaele Barretta,

S. Ali Faghidian,

Francesco Marotti de Sciarra,

Francesco Paolo Pinnola

# Timoshenko nonlocal strain gradient nanobeams: variational consistency, exact solutions and carbon nanotube Young moduli

R. Barretta [a], S. Ali Faghidian [b], F. Marotti de Sciarra [a], F. P. Pinnola [a]

[a] *Department of Structures for Engineering and Architecture, University of Naples Federico II, via Claudio 21, 80125 Naples, Italy – e-mails: rabarret@unina.it - marotti@unina.it - francescopaolo.pinnola@unina.it*

[b] *Department of Mechanical Engineering, Science and Research Branch, Islamic Azad University, Tehran, Iran – e-mail: faghidian@gmail.com*

**Abstract**

Carbon nanotubes (CNTs) are principal constituents of nanocomposites and nano-systems. CNT size-dependent response assessment is therefore a topic of current interest in Mechanics of Advanced Materials and Structures. CNTs are modelled here by a variationally consistent nonlocal strain gradient approach for Timoshenko nano-beams, extending the treatment in [Int. J. Eng. Science 143 (2019) 73-91] confined to slender structures. Scale effects are described by integral convolutions, conveniently replaced with differential and boundary nonlocal laws. The theoretical predictions, exploited to analytically estimate the reduced Young elastic modulus of CNTs, are validated by molecular dynamics simulations.

**Keywords**

Timoshenko elastic nano-beams; size effects; nonlocal integral theory; analytical modelling; CNT; reduced Young modulus.

## 1. Introduction

In contemporary decades, carbon nanotubes (CNTs) have found a variety of applications in nano-technology as fundamental elements of modern nano-devices [1, 2] and extensively exploited in nanostructure composites [3-5]. However, CNTs are well-established to exhibit size-dependent behaviors at nano-scale that cannot be modelled by the classical local theory of continuum mechanics due to the lack of material internal lengths. Conducting experiments at micro- and nano-scales is intricate in consequence of the required high-precision and normally leads to defective measurements. Development of suitable analytical and numerical models is therefore of major significance in design and optimization of small-scale devices. As a result of high computational cost of Molecular Dynamic (MD) simulations, generalized continuum mechanics strategies are usually preferred as an efficient approach in modelling nano-structural elements [6]. Rapid developments in nano-technology have caused an intensive interest in the analysis of nano-structures and carbon nanotubes since the suitable control over the exclusive properties of CNTs can potentially lead to new advancements applicable to modern nano-electro-mechanical systems (NEMS). The thematic concerning with assessment of size-dependent behavior of nanostructural composites is of current research of community of nano-engineering and -science [7-22] where some recent reviews can be found in [23, 24]. A variety of size-dependent constitutive laws are developed in the literature accounting for scale-effects in the mechanical response of nano-beams and carbon nanotubes. Nonlocal theory of elasticity formulated by Eringen [25] is perhaps one of the most popular approaches to analyze the mechanical response of nano-structures. However, such a strain-driven model leads to elastostatic problems of applicative interest that are ill-posed, due to confliction between constitutive and equilibrium boundary conditions [26]. The stress-driven nonlocal elasticity, conceived in [27], leads instead to well-posed nonlocal formulations on bounded domains and has been effectively utilized in a series of recent contributions to examine the size-dependent elastostatic [28-30], thermoelastic [31, 32] and elastodynamic [33, 34] structural problems of nano-technological interest. Nonlocal strain gradient elasticity model was introduced by Aifantis [35] by combining the Eringen nonlocal differential model and the strain gradient elasticity formulation. The thermodynamic framework for higher-order nonlocal strain gradient materials is then established by appropriate definition of the internal energy density potential [36], Reissner mixed variational functional [37] and Helmholtz free energy [38]. Application of the nonlocal strain gradient model to bounded continua, however, implies the necessity to impose additional non-classical boundary conditions. While there was a dispute in the literature on the appropriate choice of higher-order boundary conditions [39], Barretta and Marotti de Sciarra [40] provided a definite solution to this issue in the framework of modified nonlocal strain gradient theory via introducing the proper boundary conditions of constitutive type. The modified nonlocal strain gradient model has been efficiently employed to analyze the size-dependent mechanical response of elastic nano-beams [41-43]. The present study is intended to examine the size-dependent structural response and constitutive behavior of CNTs by exploiting the modified nonlocal strain gradient elasticity. A consistent variational scheme, with test fields belonging to suitably chosen functional spaces, will be utilized to introduce the modified nonlocal strain gradient elasticity, which yields well-posed mechanical formulations. Furthermore, the drawbacks of classical nonlocal strain gradient model, occurred as a result of imposing unmotivated higher-order boundary conditions, will be discussed. In particular, the structural response of carbon nanotubes is modelled by Timoshenko beam and the modified nonlocal strain gradient elasticity theory. Based on the flexural analysis of Timoshenko nano-beam, a novel reduced Euler-Young elastic modulus is introduced for CNTs and calibrated by Molecular Dynamics (MD) simulations results. New benchmark illustrations are also detected that can be profitably exploited for the design and optimization of modern NEMS. The proposed formulation of nonlocal strain gradient elasticity provides new insight on structural modelling of CNTs both from the theoretical and applicative point of view and can be extended to other technical problems of nano-mechanics.

The paper is organized as follows. Section 2 is devoted to briefly recall preliminary kinematic assumptions and differential conditions of equilibrium of Timoshenko's shear beam model. In Section 3, the nonlocal strain gradient elasticity theory for stubby beams is formulated. Static elasticity and analytical solutions are established and discussed in Sections 4 and 5. Section 6 presents a novel formulation of the reduced Euler-Young elastic modulus that can efficiently incorporate the geometrical properties of carbon nanotubes. Conclusions are finally drawn in Section 7.

## 2. Shear nano-beam model

A homogeneous isotropic nano-beam of length $L$ with cross-sectional domain $\Omega$ is considered. The $x$–, $y$– and $z$–coordinates are taken along the length, the thickness and the width of the beam, as schematically illustrated in Fig. 1. The geometry and the applied loads of the beam are such that the displacements $(s_x, s_y, s_z)$ along the axes $(x, y, z)$ are functions of the $x$– and $y$–coordinates. It is further assumed that the displacement $s_z$ is identically zero and that the nonlocal behaviour is negligible in the thickness direction.

The elasticity solution of Saint-Venant's problem [44, 45] can be applied to set forth the formulation of shear beam model, and consequently, the proposed Timoshenko nano-beam model is based on the displacement field with the following components

$$\begin{aligned} s_x(x,y) &= -y\,\varphi(x) \\ s_y(x,y) &= v(x) \\ s_z(x,y) &= 0 \end{aligned} \qquad (1)$$

where $v$ denotes the transversal displacement of a point on the mid-plane of the nano-beam and $\varphi$ is the rotation of the cross-section about the $z$–axis.

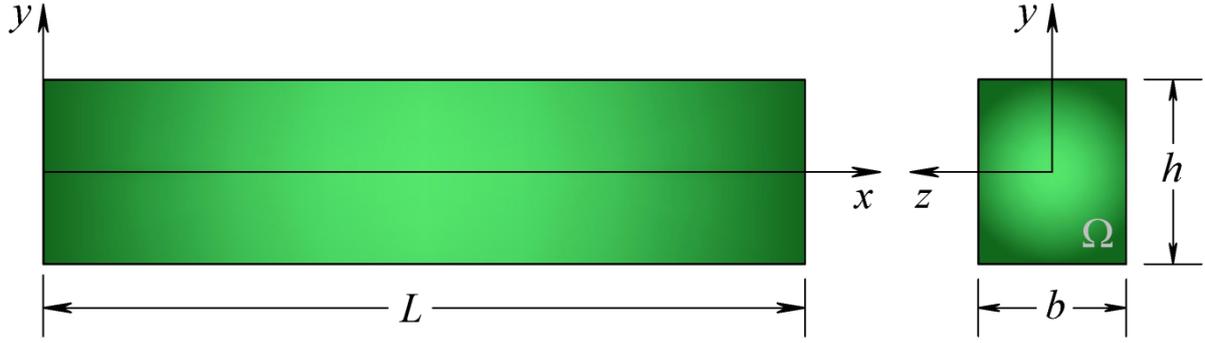

**Fig. 1**. Coordinate system and configuration of a Timoshenko nano-beam.

The only nonzero strains are the axial strain $\varepsilon_x$ and the transverse shear strain $\gamma_{xy}$ given by

$$\begin{aligned}\varepsilon_x(x,y) &= -y\,\partial_x\varphi(x) = -y\,\kappa(x) \\ \gamma_{xy}(x,y) &= \partial_x v(x) - \varphi(x) = \gamma(x)\end{aligned} \qquad (2)$$

with $\kappa$ and $\gamma$ being the flexural curvature and shear strain field, correspondingly. The symbol $\partial_x \circ$ denotes the first derivative of the function $\circ$ along the nano-beam axis $x$. Timoshenko's model requires the introduction of a shear correction factor $\chi$ to compensate the error due to the shear stress assumption and, according to Lim et al. [36], the constant value of the local Timoshenko beam theory [46] is assumed for the shear correction.

## 2.1. *Local shear model*

The axial stress $\sigma$ associated with the axial strain $\varepsilon_x$ and the shear stress $\tau$ associated with the shear strain $\gamma_{xy}$ take the form

$$\begin{aligned}\sigma(x,y) &= E\,\varepsilon_x(x,y) = -Ey\,\kappa(x) \\ \tau(x,y) &= G\,\gamma_{xy}(x,y) = G\,\gamma(x)\end{aligned} \qquad (3)$$

being $E$ and $G$ the local Euler-Young and shear moduli.

Hence, using Eq. (3), the stress resultant moment $M$ and the shear resultant $T$ are given by

$$M = -\int_\Omega \sigma y \, dA = EI\kappa(x)$$
$$T = \chi \int_\Omega \tau \, dA = \chi GA \gamma(x) \qquad (4)$$

where $\chi$ stands for shear correction factor [46]. The cross-sectional area $A$ and second moment of area $I$ about the $z$–axis are

$$(A, I) = \int_\Omega (1, y^2) \, dA \qquad (5)$$

Differential conditions of equilibrium in $[0, L]$ assume the standard form

$$\partial_x M + T = -m$$
$$\partial_x T = -q_y \qquad (6)$$

and the boundary conditions at the nano-beam end points $x = 0$ and $x = L$ are $M(0) = -\bar{M}, T(0) = -\bar{F}$ and $M(L) = \bar{M}, T(L) = \bar{F}$ where $m$ is the applied distributed bending couples system and $q_y$ is the distributed transversal loading.

## 3. Modified nonlocal strain gradient model

The elastic strain energy $U$ of Timoshenko nano-beam associated with strain gradient theory depends on elastic curvature $\kappa \in C^2([0,L]; \Re)$ and shear strain $\gamma \in C^2([0,L]; \Re)$ as [47]

$$U(\kappa, \gamma) := \frac{1}{2} \int_0^L \left( \left( K_f \kappa^2 + \ell^2 K_f (\partial_x \kappa)^2 \right) + \left( K_s \gamma^2 + \ell^2 K_s (\partial_x \gamma)^2 \right) \right) dx \qquad (7)$$

with material length-scale parameter $\ell \geq 0$ describing the influence of the higher-order strain gradient field along with $K_f$ being the local elastic bending stiffness $K_f = \int_\Omega E y^2 \, dA$ and $K_s$ is the local elastic shear stiffness $K_s = \chi \int_\Omega G \, dA$.

Motivated by the seminal approach of Eringen [25], the flexure of Timoshenko nano-beams is formulated here consistently with the modified nonlocal strain gradient theory (MNSG).

The elastic strain energy $\Pi$ of Timoshenko nano-beam in the framework of MNSG is thus established by appropriately introducing nonlocal integral convolutions in Eq. (7)

$$\Pi(\kappa,\gamma) := \frac{1}{2}\int_0^L \left( \left(K_f(\psi_0*\kappa)\kappa + \ell^2 K_f(\psi_1*\partial_x\kappa)\partial_x\kappa\right) \right.$$
$$\left. + \left(K_s(\psi_0*\gamma)\gamma + \ell^2 K_s(\psi_1*\partial_x\gamma)\partial_x\gamma\right) \right) dx \qquad (8)$$

where the nonlocal kernels $\psi_0$ and $\psi_1$ depend on two dimensionless nonlocal parameters $\lambda_0 > 0$ and $\lambda_1 > 0$. The integral convolution of a scalar source field $f$ with a scalar kernel $\psi_j$ has been denoted by

$$(\psi_j * f)(x,\lambda_j) := \int_0^L \psi_j(x-\bar{x},\lambda_j) f(\bar{x}) d\bar{x} \qquad (9)$$

with $x$ and $\bar{x}$ representing points of the beam domain $[0,L]$. The nonlocal smoothing kernels are also assumed to fulfil the positivity and parity, symmetry, normalization and limit impulsivity properties [26].

The stress field in an elastic Timoshenko nano-beam, formulated according to MNSG, are described by the bending moments field $M$ and the shear force field $T$ determined by the variational constitutive conditions

$$\langle M,\delta\kappa\rangle := \int_0^L M(x)\delta\kappa(x)dx = \langle d\Pi(\kappa,\gamma),\delta\kappa\rangle$$
$$\langle T,\delta\gamma\rangle := \int_0^L T(x)\delta\gamma(x)dx = \langle d\Pi(\kappa,\gamma),\delta\gamma\rangle \qquad (10)$$

for any virtual elastic curvature field $\delta\kappa \in C_0^1([0,L];\Re)$ and virtual shear strain field $\delta\gamma \in C_0^1([0,L];\Re)$ having compact support in $[0,L]$. Introducing the expression of $\Pi$ as Eq. (8), the directional derivatives of the elastic strain energy along a virtual elastic curvature and virtual shear strain are evaluated while applying integration by parts

$$\begin{aligned}
\langle d\Pi(\kappa,\gamma),\delta\kappa\rangle &= \int_0^L \left(K_f(\psi_0 * \kappa)\delta\kappa + \ell^2 K_f(\psi_1 * \partial_x\kappa)(\partial_x\delta\kappa)\right)dx \\
&= \int_0^L \left(K_f(\psi_0 * \kappa) - \ell^2 K_f \partial_x(\psi_1 * \partial_x\kappa)\right)\delta\kappa\, dx \\
&\quad + \ell^2 K_f \left((\psi_1 * \partial_x\kappa)\delta\kappa\big|_{x=L} - (\psi_1 * \partial_x\kappa)\delta\kappa\big|_{x=0}\right)
\end{aligned}$$

$$\begin{aligned}
\langle d\Pi(\kappa,\gamma),\delta\gamma\rangle &= \int_0^L \left(K_s(\psi_0 * \gamma)\delta\gamma + \ell^2 K_s(\psi_1 * \partial_x\gamma)(\partial_x\delta\gamma)\right)dx \\
&= \int_0^L \left(K_s(\psi_0 * \gamma) - \ell^2 K_s \partial_x(\psi_1 * \partial_x\gamma)\right)\delta\gamma\, dx \\
&\quad + \ell^2 K_s \left((\psi_1 * \partial_x\gamma)\delta\gamma\big|_{x=L} - (\psi_1 * \partial_x\gamma)\delta\gamma\big|_{x=0}\right)
\end{aligned}$$

(11)

where the linear independency of virtual elastic curvature and virtual shear strain fields $\delta\kappa, \delta\gamma$ is implemented. Both test fields $\delta\kappa \in C_0^1([0,L];\Re)$ and $\delta\gamma \in C_0^1([0,L];\Re)$ in the variational constitutive conditions Eq. (10) have compact supports, so that $\delta\kappa\big|_{x=L} = \delta\kappa\big|_{x=0} = 0$ and $\delta\gamma\big|_{x=L} = \delta\gamma\big|_{x=0} = 0$, and therefore, the boundary terms in Eq. (11) are disappearing.

Employing a standard localization procedure, bending moment *M* and shear force *T* can be respectively detected in terms of elastic curvature $\kappa$ and shear strain field $\gamma$ while imposing the variational condition Eq. (10)

$$\begin{aligned}
M &= K_f(\psi_0 * \kappa) - \ell^2 K_f \partial_x(\psi_1 * \partial_x\kappa) \\
T &= K_s(\psi_0 * \gamma) - \ell^2 K_s \partial_x(\psi_1 * \partial_x\gamma)
\end{aligned}$$

(12)

Utilizing the definition of integral convolution along with the kinematic compatibility conditions, the variationally consistent nonlocal strain gradient model for Timoshenko nano-beams is formulated by expressing the bending moment *M* and the shear force *T* as follows

$$\begin{aligned}
M &= K_f(\psi_0 * \partial_x\varphi)(x,\lambda_0) - \ell^2 K_f \partial_x(\psi_1 * \partial_x^2\varphi)(x,\lambda_1) \\
&= K_f \int_0^L \psi_0(x-\bar{x},\lambda_0)\partial_{\bar{x}}\varphi(\bar{x})d\bar{x} - \ell^2 K_f \partial_x \int_0^L \psi_1(x-\bar{x},\lambda_1)\partial_{\bar{x}}^2\varphi(\bar{x})d\bar{x} \\
T &= K_s(\psi_0 * \gamma)(x,\lambda_0) - \ell^2 K_s \partial_x(\psi_1 * \partial_x\gamma)(x,\lambda_1) \\
&= K_s \int_0^L \psi_0(x-\bar{x},\lambda_0)\gamma(\bar{x})d\bar{x} - \ell^2 K_s \partial_x \int_0^L \psi_1(x-\bar{x},\lambda_1)\partial_{\bar{x}}\gamma(\bar{x})d\bar{x}
\end{aligned}$$

(13)

Following Lim et al. [36], we assume that the nonlocal parameters are coincident, i.e. $\lambda := \lambda_0 = \lambda_1$, and that the kernels $\psi_0$ and $\psi_1$ coincide with the bi-exponential averaging function below, widely adopted in applications of Engineering Science

$$\psi(x, L_c) = \frac{1}{2L_c} e^{-\frac{|x|}{L_c}} \tag{14}$$

being $L_c = \lambda L$ the characteristic length of Eringen nonlocal elasticity. The bi-exponential function fulfils positivity, symmetry, normalization and impulsivity, see e.g. [26].

Following Barretta and Marotti de Sciarra [40], it can be proved that the nonlocal strain gradient integral relations Eq. (13) for Timoshenko nano-beams are equivalent to differential laws subjected to (non-classical) boundary conditions as reported in the next Proposition.

**Proposition 3.1 (Constitutive equivalence property).**

*The nonlocal strain gradient constitutive laws for Timoshenko nano-beams Eq. (13), equipped with the bi-exponential kernel Eq. (14) with $x \in [0, L]$, are equivalent to the differential relations*

$$\begin{aligned}
K_f \partial_x \varphi(x) - \ell^2 K_f \partial_x^3 \varphi(x) &= M(x, L_c, \ell) - L_c^2 \partial_x^2 M(x, L_c, \ell) \\
K_s \gamma(x) - \ell^2 K_s \partial_x^2 \gamma(x) &= T(x, L_c, \ell) - L_c^2 \partial_x^2 T(x, L_c, \ell)
\end{aligned} \tag{15}$$

*subjected to the following four constitutive boundary conditions (CBCs)*

$$\begin{aligned}
\partial_x M(0, L_c, \ell) &= \frac{1}{L_c} M(0, L_c, \ell) + \frac{\ell^2}{L_c^2} K_f \partial_x^2 \varphi(0) \\
\partial_x M(L, L_c, \ell) &= -\frac{1}{L_c} M(L, L_c, \ell) + \frac{\ell^2}{L_c^2} K_f \partial_x^2 \varphi(L) \\
\partial_x T(0, L_c, \ell) &= \frac{1}{L_c} T(0, L_c, \ell) + \frac{\ell^2}{L_c^2} K_s \partial_x \gamma(0) \\
\partial_x T(L, L_c, \ell) &= -\frac{1}{L_c} T(L, L_c, \ell) + \frac{\ell^2}{L_c^2} K_s \partial_x \gamma(L)
\end{aligned} \tag{16}$$

## 4. Static elastic model

The expression of the bending moment $M$ can be obtained from Eq. (15)$_1$ employing the differential conditions of equilibrium Eq. (6) to get

$$M(x, L_c, l) = K_f \partial_x \varphi(x) - \ell^2 K_f \partial_x^3 \varphi(x) + L_c^2 (q_y(x) - \partial_x m(x)) \tag{17}$$

Analogously, using the differential condition of equilibrium Eq. (6)$_2$, the expression of the shear force $T$ follows from Eq. (15)$_2$ to get

$$T(x, L_c, l) = K_s \gamma(x) - \ell^2 K_s \partial_x^2 \gamma(x) - L_c^2 \partial_x q_y(x) \tag{18}$$

The nonlocal differential equilibrium equations for the considered Timoshenko nano-beam according to MNSG can be obtained by differentiating Eq. (15) with respect to $x$ and then by summing up the derivative of Eq. (15)$_1$ to Eq. (15)$_2$. Hence, recalling the differential conditions of equilibrium Eq. (6), the nonlocal differential equilibrium equations for the Timoshenko nano-beam consistent with MNSG are

$$\begin{aligned} K_f \partial_x^2 \varphi(x) - \ell^2 K_f \partial_x^4 \varphi(x) + K_s \gamma(x) - \ell^2 K_s \partial_x^2 \gamma(x) = -m(x) + L_c^2 \partial_x^2 m(x) \\ K_s \partial_x \gamma(x) - \ell^2 K_s \partial_x^3 \gamma(x) = -q_y(x) + L_c^2 \partial_x^2 q_y(x) \end{aligned} \tag{19}$$

equipped with the classical boundary conditions at the nano-beam end points $x = 0$ and $x = L$ and the constitutive boundary conditions Eq. (16).

**Remark 4.1** The nonlocal differential equilibrium equations Eq. (19) for the MNSG Timoshenko nano-beam can be compared to the corresponding ones pertaining to the Timoshenko nano-beam associated with strain gradient elasticity provided by Marotti de Sciarra and Barretta [47]. It is worth noting that assuming that the gradient parameters $b$ and $c$ in [47] as $b = c = \ell$ while $L_c \to 0^+$, the differential conditions of equilibrium of Timoshenko nano-beam associated with strain gradient elasticity can be obtained via simplifying Eq. (19) as

$$K_f \partial_x^2 \varphi(x) - \ell^2 K_f \partial_x^4 \varphi(x) + K_s \gamma(x) - \ell^2 K_s \partial_x^2 \gamma(x) = -m(x)$$
$$K_s \partial_x \gamma(x) - \ell^2 K_s \partial_x^3 \gamma(x) = -q_y(x) \qquad (20)$$

Additionally, the higher-order boundary conditions in [47] can be also detected via simplifying MNSG CBC Eq. (16) on beam ends as

$$\ell^2 K_f \partial_x^2 \varphi(0) = 0$$
$$\ell^2 K_f \partial_x^2 \varphi(L) = 0$$
$$\ell^2 K_s \partial_x \gamma(0) = 0 \qquad (21)$$
$$\ell^2 K_s \partial_x \gamma(L) = 0$$

The gradient constitutive law and higher-order boundary conditions of the well-established strain gradient model of Timoshenko elastic beams can be successfully recovered for vanishing nonlocal characteristic parameter.

## 5. Analytical solutions

In the present study, we consider a nano-cantilever subject to a transverse applied force $\bar{F}$ at the free end point and a simply supported nano-beam subject to a uniform load $q_y(x) = q$. The influences of nonlocal parameter $L_c$, material length-scale parameter $\ell$, slenderness ratio $L/h$ and shear deformation on the bending behavior are examined.

### 5.1. *Nano-cantilever*

Let us consider a nano-cantilever with length $L$ subject to a transverse applied force $\bar{F}$ at the free end point. The bending and shear stiffnesses are assumed to be constant along the nano-cantilever, so that the differential equations Eq. (19) of the MNSG Timoshenko beam become

$$K_f \partial_x^2 \varphi(x) - \ell^2 K_f \partial_x^4 \varphi(x) + K_s \gamma(x) - \ell^2 K_s \partial_x^2 \gamma(x) = 0$$
$$\partial_x \gamma(x) - \ell^2 \partial_x^3 \gamma(x) = 0 \qquad (22)$$

equipped with the set of boundary conditions

$$v(0) = 0$$
$$\varphi(0) = 0$$
$$M(L) = 0$$
$$T(L) = \bar{F}$$

$$\partial_x M(0, L_c, \ell) = \frac{1}{L_c} M(0, L_c, \ell) + \frac{\ell^2}{L_c^2} K_f \partial_x^2 \varphi(0)$$

$$\partial_x T(0, L_c, \ell) = \frac{1}{L_c} T(0, L_c, \ell) + \frac{\ell^2}{L_c^2} K_s \partial_x \gamma(0) \qquad (23)$$

$$\partial_x M(L, L_c, \ell) = -\frac{1}{L_c} M(L, L_c, \ell) + \frac{\ell^2}{L_c^2} K_f \partial_x^2 \varphi(L)$$

$$\partial_x T(L, L_c, \ell) = -\frac{1}{L_c} T(L, L_c, \ell) + \frac{\ell^2}{L_c^2} K_s \partial_x \gamma(L)$$

where shear force and bending moment fields are

$$T(x, L_c, l) = K_s \gamma(x) - \ell^2 K_s \partial_x^2 \gamma(x)$$
$$M(x, L_c, l) = K_f \partial_x \varphi(x) - \ell^2 K_f \partial_x^3 \varphi(x) \qquad (24)$$

Note that Eq. (23)$_{1-4}$ are the classical boundary conditions of the cantilever subjected to a transverse applied force $\bar{F}$ at $x = L$. Eq. (23)$_{5-8}$ are the constitutive boundary conditions associated with the MNSGT Timoshenko beam. It is apparent that no higher-order boundary conditions have to be added to solve the nonlocal problem and the CBCs are consistently associated with the nonlocal strain gradient integral model.

The analytical solution of Eq. (22)$_2$ of the MNSG Timoshenko beam can be expressed as

$$\gamma(x) = e^{\frac{x}{\ell}} \ell A_1 - e^{-\frac{x}{\ell}} \ell A_2 + A_3 \qquad (25)$$

Hence, the nonlocal Eq. (22)$_1$ of the MNSG Timoshenko beam becomes

$$K_f \partial_x^2 \varphi(x) - \ell^2 K_f \partial_x^4 \varphi(x) = -K_s \gamma(x) + \ell^2 K_s \partial_x^2 \gamma(x) \qquad (26)$$

and the rotation field is

$$\varphi(x) = -\frac{K_s}{2K_f} A_3 x^2 + e^{\frac{x}{\ell}} \ell^2 A_4 + e^{-\frac{x}{\ell}} \ell^2 A_5 + A_6 + A_7 x \tag{27}$$

Finally, the transversal displacement is recovered by means of the kinematic compatibility relation Eq. (2)₂, i.e. $\gamma = \partial_x v - \varphi$, so that we have

$$v(x) = -\frac{K_s}{6K_f} A_3 x^3 + A_3 x + e^{\frac{x}{\ell}} \ell^2 (A_1 + \ell A_4) - e^{-\frac{x}{\ell}} \ell^2 (-A_2 + \ell A_5) + A_6 x + \frac{1}{2} A_7 x^2 + A_8 \tag{28}$$

where $A_j$, with $j = \{1,...,8\}$, are the unknown constants to be determined by the well-known classical boundary conditions and the constitutive boundary conditions reported in Eq. (23).

For illustrative purpose, dimensionless transversal displacements and rotations are defined by

$$v^* = v \frac{K_f}{\overline{F}L^3}, \qquad \varphi^* = \varphi \frac{K_f}{\overline{F}L^2} \tag{29}$$

For suitable comparison to the counterpart results available in the literature, the beam is assumed to have rectangular cross-section with width $b$ and thickness $h$. The thickness of the nano-cantilever is taken as $h$=1 nm and the Poisson's ratio is $0.3$. The slenderness ratio $L/h$ is also assumed in the range $\{10, 20, 50\}$.

Tables 1, 2 and 3 present the dimensionless maximum deflections for different values of the nonlocal parameter $L_c$, material length-scale parameter $\ell$ and slenderness ratio $L/h$ based on the proposed modified nonlocal strain gradient model for Timoshenko and Bernoulli–Euler nano-beams. The non-dimensional maximum deflections of Bernoulli–Euler nano-beams in the framework of MNSG are also re-developed here following the analytical approach of Apuzzo et al. [42]. In the sequel, the acronyms NSG and MNSG, respectively, denote the nonlocal strain gradient model and modified nonlocal strain gradient theory. Besides, the acronyms BEM and TBM stand for Bernoulli–Euler beam model and Timoshenko beam model, correspondingly.

Fig. 2 illustrates the 3D plot of variation of dimensionless transversal displacement of the nano-cantilever subjected to transversal force $\bar{F}$ versus the nonlocal parameter $L_c$ and the material length-scale parameter $\ell$. Both the nonlocal and gradient length-scale parameters are ranging in the interval $]0,2[$ nm and the slenderness ratio of the nano-beam is assumed as $L/h = 10$ in Fig. 2.

It is inferred from Fig. 2 that for a given value of the material length-scale parameter $\ell$, the increase of the nonlocal parameter $L_c$ has the effect of softening the flexural response of MNSG nano-cantilever with respect to the local beam model since the dimensionless transversal displacements increase. The flexural displacement field of nano-beam decreases as the material length-scale parameter $\ell$ increases, and therefore, the adopted modified nonlocal strain gradient model exhibits a stiffening behavior in terms of material length-scale parameter $\ell$ for a given value of nonlocal parameter $L_c$. As the nonlocal and gradient length-scale parameters approach zero $L_c, \ell \to 0^+$, the elastostatic flexure of nano-beam associated with the modified nonlocal strain gradient model is coincident with the transversal displacements of local elastic beam models.

The effect of slenderness ratio on the dimensionless maximum deflections is depicted in Fig. 3 for different values of nonlocal parameter $L_c$ ranging in the set $\{0^+, 0.5, 1.0, 1.5, 2.0\}$ nm and material length-scale parameter $\ell = 1$ nm. As expected, it can be observed from Fig. 3 that the dimensionless maximum deflections increase as the nonlocal parameter increases. While the difference between Timoshenko and Bernoulli-Euler nano-beams model tend to vanish for increasing values of the slenderness ratio $L/h$, shear effects are significant for stubby beams. Moreover, the flexural response of Bernoulli-Euler nano-beams underestimates the counterpart results of Timoshenko nano-beam model.

According to Fig. 3, it is of interest that the dimensionless maximum deflections tend to increase as the slenderness ratio increases when $L_c \to 0^+$ but decrease with the increase of slenderness ratio for non-vanishing nonlocal parameter $L_c$. This is due to the dominant stiffness-softening effects of the nonlocal parameter $L_c$ for tip-loaded nano-cantilevers.

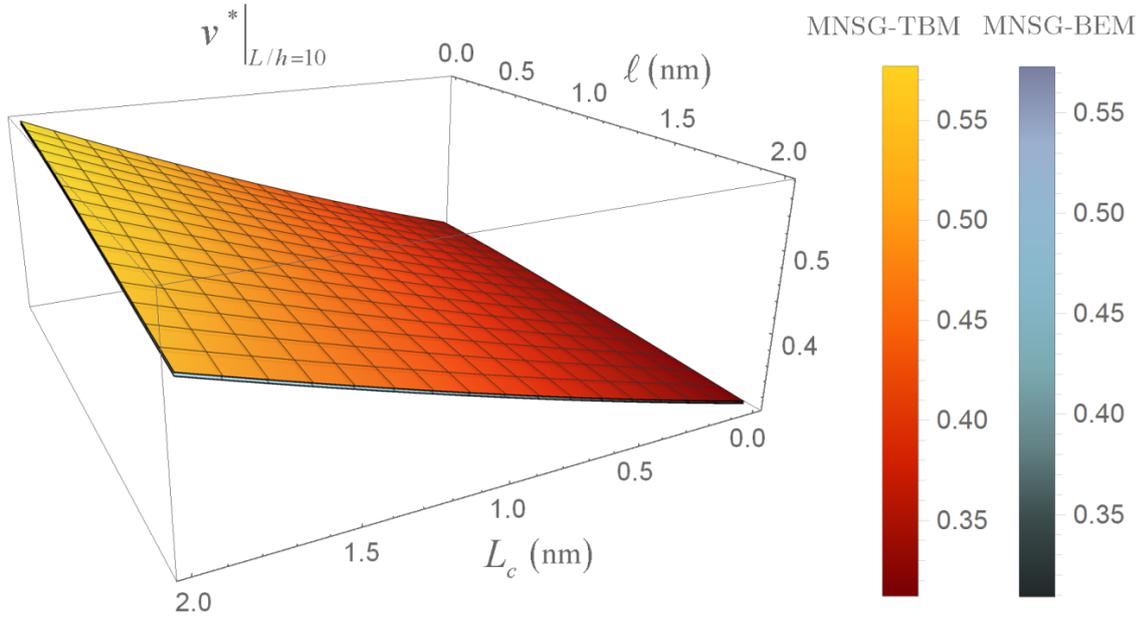

**Fig. 2.** Nano-cantilever: dimensionless maximum deflection $v^*$ vs. $L_c$ and $\ell$ for $L/h = 10$

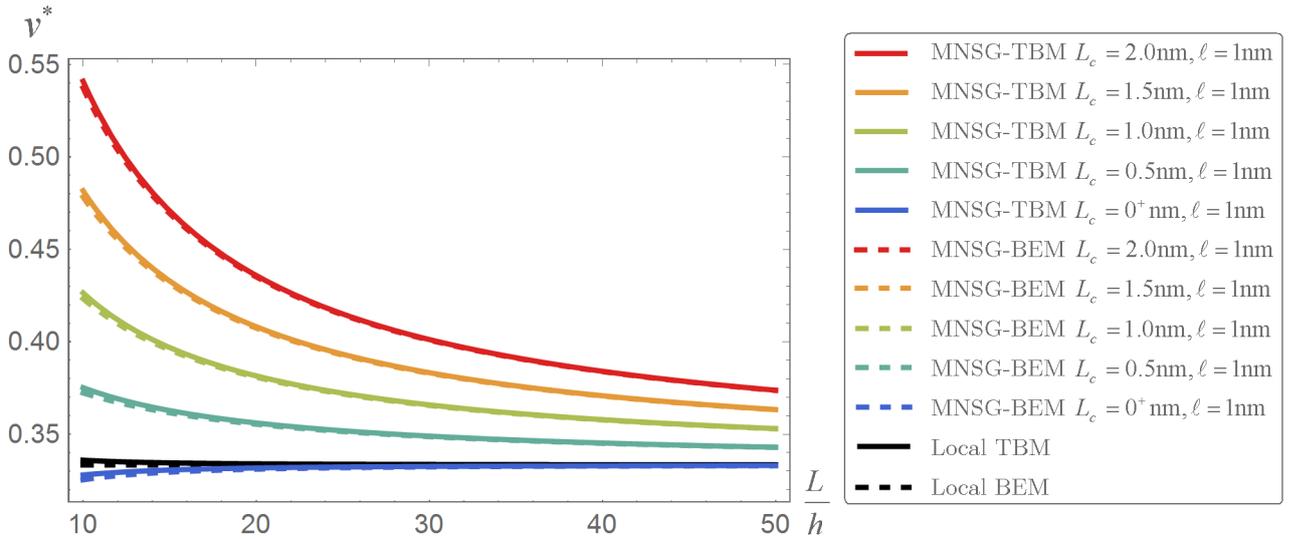

**Fig. 3.** Nano-cantilever: effects of the slenderness ratio $L/h$ on the dimensionless maximum deflection

## 5.2 *Simply supported nano-beam*

Let us consider a simply supported nano-beam with length $L$ subjected to a transverse uniform load $q$. Bending and shear stiffnesses are assumed to be constant along the nano-beam, so that the differential equations Eq. (19) of the MNSG Timoshenko beam become

$$K_f \partial_x^2 \varphi(x) - \ell^2 K_f \partial_x^4 \varphi(x) + K_s \gamma(x) - \ell^2 K_s \partial_x^2 \gamma(x) = 0$$
$$K_s \partial_x \gamma(x) - \ell^2 K_s \partial_x^3 \gamma(x) = -q$$
(30)

and the boundary conditions are

$$v(0) = 0$$
$$M(0) = 0$$
$$v(L) = 0$$
$$M(L) = 0$$
$$\partial_x M(0, L_c, \ell) = \frac{1}{L_c} M(0, L_c, \ell) + \frac{\ell^2}{L_c^2} K_f \partial_x^2 \varphi(0)$$
$$\partial_x T(0, L_c, \ell) = \frac{1}{L_c} T(0, L_c, \ell) + \frac{\ell^2}{L_c^2} K_s \partial_x \gamma(0)$$
$$\partial_x M(L, L_c, \ell) = -\frac{1}{L_c} M(L, L_c, \ell) + \frac{\ell^2}{L_c^2} K_f \partial_x^2 \varphi(L)$$
$$\partial_x T(L, L_c, \ell) = -\frac{1}{L_c} T(L, L_c, \ell) + \frac{\ell^2}{L_c^2} K_s \partial_x \gamma(L)$$
(31)

being the bending moment and the shear force expressed by

$$T(x, L_c, l) = K_s \gamma(x) - \ell^2 K_s \partial_x^2 \gamma(x)$$
$$M(x, L_c, l) = K_f \partial_x \varphi(x) - \ell^2 K_f \partial_x^3 \varphi(x) + L_c^2 q$$
(32)

Eq. (31)$_{1-4}$ are the classical boundary conditions of a simply supported beam subjected to a transverse uniform loading and Eq. (31)$_{5-8}$ represent the constitutive boundary conditions associated with the MNSG Timoshenko beam. It is apparent that no higher-order boundary conditions have to be added to solve the nonlocal strain gradient problem.

Following a procedure analogous to the one shown in the previous example, the solution of the nonlocal strain gradient equation (30)$_2$ of the MNSG Timoshenko beam is given by

$$\gamma(x) = e^{\frac{x}{\ell}} \ell A_1 - e^{-\frac{x}{\ell}} \ell A_2 + A_3 - \frac{q}{K_s} x \tag{33}$$

Hence, the nonlocal strain gradient Eq. (30)$_1$ of the MNSG Timoshenko beam becomes

$$K_f \partial_x^2 \varphi(x) - \ell^2 K_f \partial_x^4 \varphi(x) = -K_s \gamma(x) + \ell^2 K_s \partial_x^2 \gamma(x) \tag{34}$$

and the rotation field is

$$\varphi(x) = -\frac{K_s}{2K_f} A_3 x^2 + e^{\frac{x}{\ell}} \ell^2 A_4 + e^{-\frac{x}{\ell}} \ell^2 A_5 + A_6 + A_7 x + \frac{q}{6K_f} x^3 \tag{35}$$

Finally, the transversal displacement is recovered by means of the kinematic compatibility relation Eq. (2)$_2$, i.e. $\gamma = \partial_x v - \varphi$, so that we have

$$v(x) = -\frac{K_s}{6K_f} A_3 x^3 + A_3 x + e^{\frac{x}{\ell}} \ell^2 (A_1 + \ell A_4) - e^{-\frac{x}{\ell}} \ell^2 (-A_2 + \ell A_5) + A_6 x + \frac{1}{2} A_7 x^2 + A_8 \\ - \frac{q}{2K_s} x^2 + \frac{q}{24K_f} x^4 \tag{36}$$

where $A_j$, with $j = \{1,...,8\}$, are the unknown integration constants that can be determined by the well-known classical boundary conditions and the constitutive boundary conditions reported in Eq. (31).

For illustrative purpose, dimensionless transversal displacements and rotations are defined by

$$v^* = v \frac{100 K_f}{qL^4}, \qquad \varphi^* = \varphi \frac{100 K_f}{qL^3} \tag{37}$$

In order to compare the outcomes with the pertinent results available in the literature, the beam is assumed to have rectangular cross-section with width $b$ and thickness $h$.

The thickness of the nano-cantilever is taken as $h$=1 nm and the Poisson's ratio is 0.3. The slenderness ratio $L/h$ is also assumed to be included in the set $\{10, 20, 50\}$.

Tables 4, 5 and 6 exhibit the dimensionless maximum deflections for different values of the nonlocal parameter $L_c$, material length-scale parameter $\ell$ and slenderness ratio $L/h$ based on the proposed modified nonlocal strain gradient model (MNSG) in comparison to the results obtained by the nonlocal strain gradient model (NSG) reported in Lu et al. [48] for both Timoshenko and Bernoulli–Euler beam models.

Transversal displacements are evaluated in Lu et al. [48] by various higher-order shear deformation theories on the basis of the nonlocal strain gradient model of Lim et al. [36], so that non-classical higher-order boundary conditions have to be added. Accordingly, the higher-order boundary conditions of the strain gradient model of Timoshenko elastic beam as Eq. (21) are applied in Lu et al. [48] implicating vanishing of derivative of the flexural curvature and derivative of the shear strain field at the end cross-sections. The prescription of unmotivated higher-order boundary conditions as Eq. (21) will lead to over-constrained [39] and inconsistent boundary value problem [40] in the nonlocal strain gradient theory.

The 3D plot of variation of the dimensionless transversal displacement of a simply supported nano-beam subjected to a uniform loading in terms of nonlocal parameter $L_c$ and material length-scale parameter $\ell$ is depicted in Fig. 4. The small-scale characteristic parameters and the slenderness ratio of the nano-beam are assumed to have the same ranging set as Fig. 2. It is deduced form the illustrative results that the dimensionless transversal displacements of nano-beam predicted by the MNSG increase as the nonlocal parameter $L_c$ increases, and accordingly, exhibit a softening behavior in terms of the nonlocal parameter $L_c$ for a given value of the material length-scale parameter $\ell$.

The modified nonlocal strain gradient theory demonstrates a stiffening behaviour in terms of the material length-scale parameter $\ell$, that is a larger $\ell$ involves a smaller transversal displacement for a given value of nonlocal parameter $L_c$. The transversal displacement of the local beam can be also recovered as the nonlocal and gradient length-scale parameters tend to zero $L_c, \ell \to 0^+$. Examination of the flexural response of simply supported nano-beams subjected to uniform transverse loading as detected in Lu et al. [48] and recalled in Tables 4-6 reveals a controversial drawback of NSG. The maximum deflection of NSG nano-beams with simply supported ends exhibit a peculiar response to coincide with the flexural response of local elastic beam as the small-scale characteristic parameters are identical. Such a peculiar structural response is absent in flexural results associated with the modified nonlocal strain gradient model which yields thus consistent size-dependent responses.

In Fig. 5, the effect of the slenderness ratio on the transversal displacement of nano-beam is demonstrated for simply support boundary conditions. The small-scale characteristic parameters and the slenderness ratio of the nano-beam are assumed to have the same ranging set as in Fig. 3. While the effects of transverse shear deformation are notable for small values of slenderness ratio $L/h$, the discrepancy between the flexural responses of Timoshenko and Bernoulli-Euler nano-beams tends to vanish for increasing values of the slenderness ratio. Furthermore, transversal displacements of Bernoulli-Euler nano-beams underestimate the flexural response of Timoshenko nano-beam model.

In accordance with Fig. 5, it can be also observed that the dimensionless maximum deflections decrease with the increase of slenderness ratio when $L_c \geq \ell$ and tend to increase as the slenderness ratio increases when $L_c < \ell$. This is due to the fact that when the material length-scale parameter is smaller than or equal to the nonlocal parameter, the nonlocal effect plays a dominant role, which makes the nano-beam to exhibit a stiffness-softening effect.

Finally, the dimensionless maximum transversal deflections of the MNSG Timoshenko beam tend to the ones of the local Timoshenko model with the increasing of the nano-beam length since, as well-known, size-effects are more prominent for stubby nano-beams.

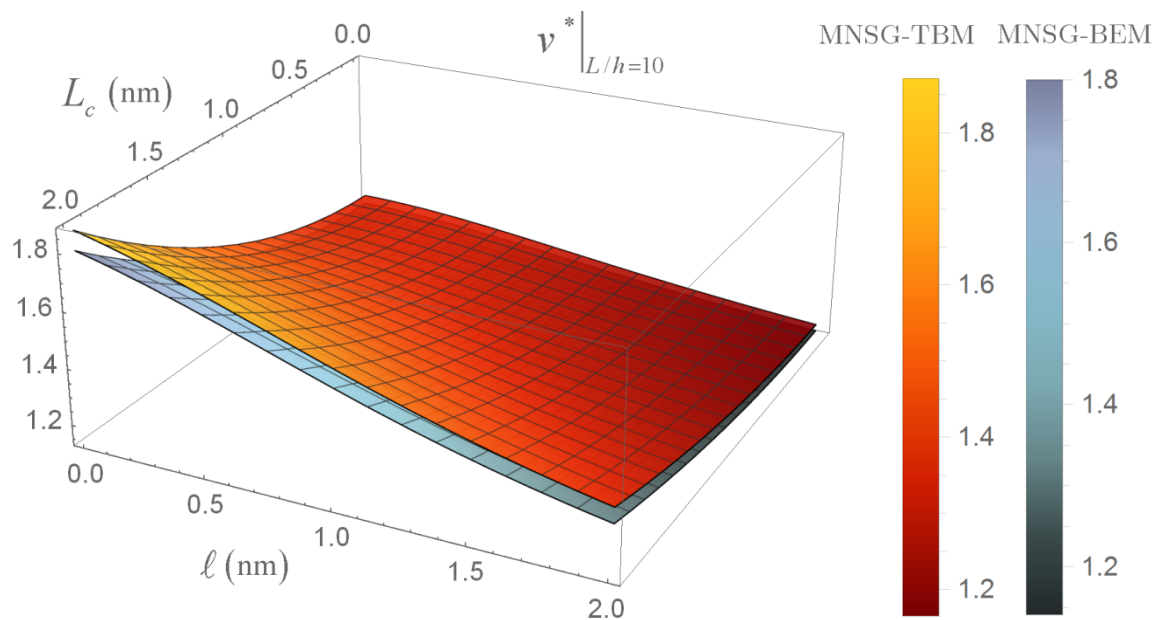

**Fig. 4.** Simply supported nano-beam: dimensionless maximum deflection $v^*$ vs. $L_c$ and $\ell$ for

$L/h = 10$

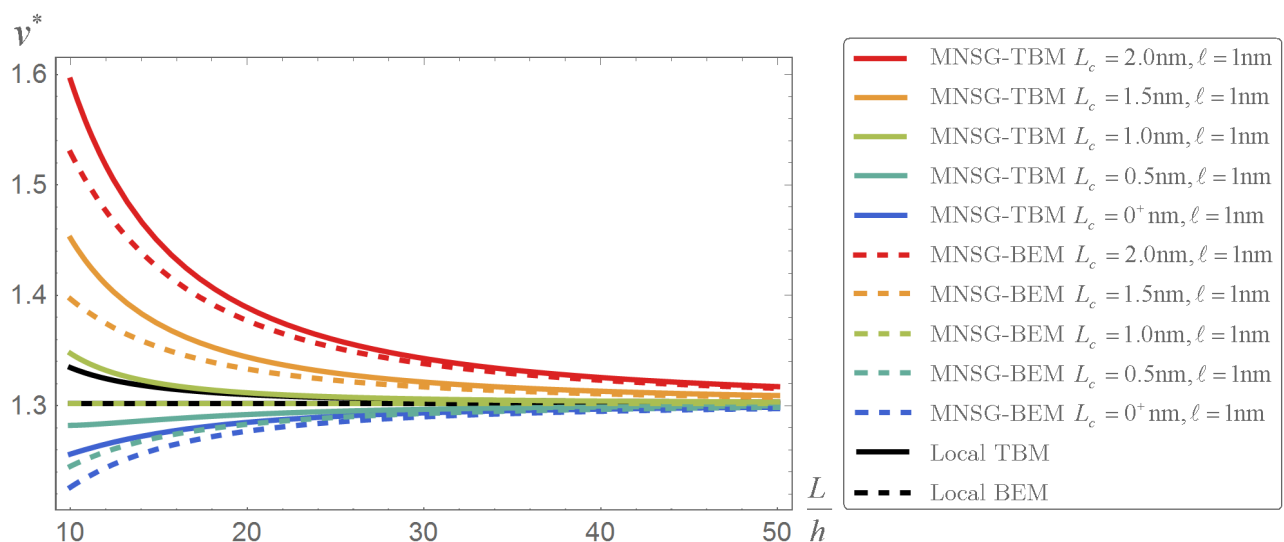

**Fig. 5.** Simply supported nano-beam: effects of the slenderness ratio $L/h$ on the

dimensionless maximum deflection

## 6. Reduced Euler-Young elastic modulus

While the cross-section is arbitrary in the proposed MNSG Timoshenko beam model, to examine the reduced Euler-Young elastic modulus of carbon nanotubes, the CNT is treated as a thin-shell beam [6] whose cross-section is depicted in Fig. 6. So, the cross-sectional area is $A = \pi D t$ and the second moment of area $I$ about the $z$–axis can be given by $I = \pi D^3 t / 8$ with $D$ and $t$ being the diameter and thickness of the CNT, respectively.

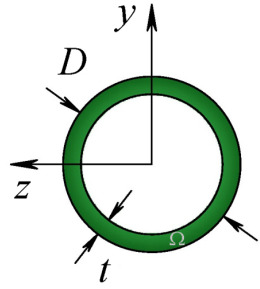

**Fig. 6**. Thin-shell cross-section exploited for elastic modulus assessment of CNT.

Determination of elastic modulus of CNT as essential constituents of advanced NEMS is of great importance. Accordingly, from Eq. (28), the maximum transversal displacement of tip loaded nano-cantilever is first evaluated as

$$v(L) = \frac{\bar{F}}{K_s}(L + 2L_c) + \frac{\bar{F}L}{3K_f}\left(-3\ell^2 + L^2 + 3L_c(L + L_c)\right) \\ + \frac{\bar{F}\ell}{K_f}\left(2\ell^2 - L_c(L + 2L_c)\right)\tanh\left(\frac{L}{2\ell}\right) \tag{38}$$

which depends on both elastic bending stiffness $K_f = \int_\Omega E y^2 dA$ and elastic shear stiffness $K_s = \chi \int_\Omega G dA$. To be compatible with the kinematic assumptions of the Timoshenko beam model, the Poisson ratio is assumed to vanish; consequently $G = E/2$ and $\chi = 1/2$ [46].

For uniform nano-beam, the reduced flexural rigidity is evaluated in view of Eq. (38) as

$$K_r = \frac{\bar{F}L^3}{v(L)}\left(\frac{1}{3} + \frac{1}{L^2}\frac{K_f}{K_s}\right)$$

$$= \frac{K_f L\left(3D^2 + 2L^2\right)}{6L_c^2 L + 6L_c\left(D^2 + L^2\right) + L\left(3D^2 - 6\ell^2 + 2L^2\right) - 6\ell\left(-2\ell^2 + L_c\left(2L_c + L\right)\right)\tanh\left(\frac{L}{2\ell}\right)} \quad (39)$$

Noticeably, the local bending stiffness $K_f$ can be recovered as the nonlocal and gradient length-scale parameters tend to zero $L_c, \ell \to 0^+$.

The maximum transversal displacement of the MNSG Timoshenko beam can be evaluated adopting the flexure analysis of local beams while utilizing the reduced flexural rigidity as Eq. (39). Assuming constant elastic and shear moduli, the elastic bending stiffness is $K_f = EI$. Accordingly, the reduced Euler-Young elastic modulus $E_r$ can be detected as

$$E_r = \frac{EL\left(3D^2 + 2L^2\right)}{6L_c^2 L + 6L_c\left(D^2 + L^2\right) + L\left(3D^2 - 6\ell^2 + 2L^2\right) - 6\ell\left(-2\ell^2 + L_c\left(2L_c + L\right)\right)\tanh\left(\frac{L}{2\ell}\right)} \quad (40)$$

The introduced reduced Euler-Young elastic modulus $E_r$ is not only dependent on the length of CNTs but also on the diameter of CNTs, and consequently, can effectively incorporate the geometrical properties of carbon nanotubes.

It is well-established in the sense of the theory of elasticity that the Euler-Young elastic modulus is a material-dependent constitutive parameter. However, the classical elasticity is confined to analyse structures that are large in comparison with characteristic length-scales of the continuum. When the principal feature of interest is the nanoscopic analysis of the field quantities in nano-structure, the classical local elasticity provides outcomes of limited technical interest. The proposed modified nonlocal strain gradient theory not only can efficiently capture the nonlocality via nonlocal convolution integrals but also can effectively incorporate the significance of strain gradient fields. The established reduced Euler-Young elastic modulus $E_r$ based on the MNSG depends on both the nonlocal parameter $L_c$ and the material length-scale parameter $\ell$. Therefore, it can be successfully calibrated with the experimental results or molecular dynamic simulations.

To study the size-dependent response of CNTs, a single-walled carbon nanotube of armchair $(10,10)$ is selected here. The diameter $D$ of the single-walled CNT $(n,m)$ is well-known to be $D = a\sqrt{3(n^2 + mn + m^2)}/\pi$ with the carbon-carbon bond length $a = 0.142$ nm [6]. Therefore, $D = 1.356$ nm for a single-walled CNT of armchair $(10,10)$ with the effective thickness of $t = 0.34$ nm. The results of elastic modulus predicted by MD simulations as reported in [49, 50] with the local elastic modulus $E = 909.5$ GPa will be exploited. Accordingly, detection of the nonlocal and gradient characteristic parameters is based on best agreement with MD simulations results.

In Fig. 7, the proposed reduced Euler-Young elastic modulus $E_r$ is plotted together with the MD simulation results [50] versus the length of the single-walled CNT. An excellent agreement can be reached between the proposed reduced Euler-Young elastic modulus based on MNSG and MD simulation results setting the nonlocal parameter $L_c = 0.0409797$ nm and the material length-scale parameter $\ell = 0.0000274$ nm. As a result of dominant small-scale effects of the nonlocal parameter for MNSG tip-loaded nano-cantilevers, the detected value of the nonlocal parameter is larger than the material length-scale parameter. The illustrated results in Fig. 7 are also compared with the reduced Euler-Young elastic modulus determined based on the MNSG elastic rod model as $\bar{E}_r = EL/(L + 2L_c)$ with $L_c = 0.0534$ nm [43]. Noticeably, the introduced reduced Euler-Young elastic modulus $E_r$ not only comprises both nonlocal and gradient characteristic parameters but also can effectively integrate the geometrical parameters of CNTs. Lastly, reduced Euler-Young elastic modulus approaches the local elastic modulus as the length of the single-walled CNT increases, due to the dominance of small-scale effects in stubby nano-beams.

To further examine the reduced Euler-Young elastic modulus, the 3D plot of $E_r$ based on the proposed MNSG Timoshenko beam model versus the nonlocal parameter $L_c$ and the material length-scale parameter $\ell$ is given in Fig. 8.

Both nonlocal and gradient length-scale parameters are ranging in the interval $]0,2[$ nm and two values for the length of the nano-beam are assumed as $L = 5$ nm and $L = 10$ nm in Fig. 8.

In accordance with Fig. 8, the reduced Euler-Young elastic modulus noticeably illustrates a stiffness-hardening behaviour in terms of the material length-scale parameter $\ell$ and stiffness-softening response in terms of nonlocal parameter $L_c$ in comparison with local elastic modulus. As the nonlocal and gradient length-scale parameters approach zero $L_c, \ell \to 0^+$, the reduced Euler-Young elastic modulus $E_r$ coincides with the local elastic modulus $E$. It is also of interest that the reduced Euler-Young elastic modulus tends to decrease as the length of CNT increases when $L_c \to 0^+$ but rapidly increase with the increase of the length of CNT for non-vanishing nonlocal parameter $L_c$, once more, as a result of influential size-dependent effects of the nonlocal parameter $L_c$.

The numerical values of reduced Euler-Young elastic modulus for single-walled CNT of armchair $(10,10)$ for different values of the nonlocal parameter $L_c$, material length-scale parameter $\ell$ and length of CNT are collected in Table 7.

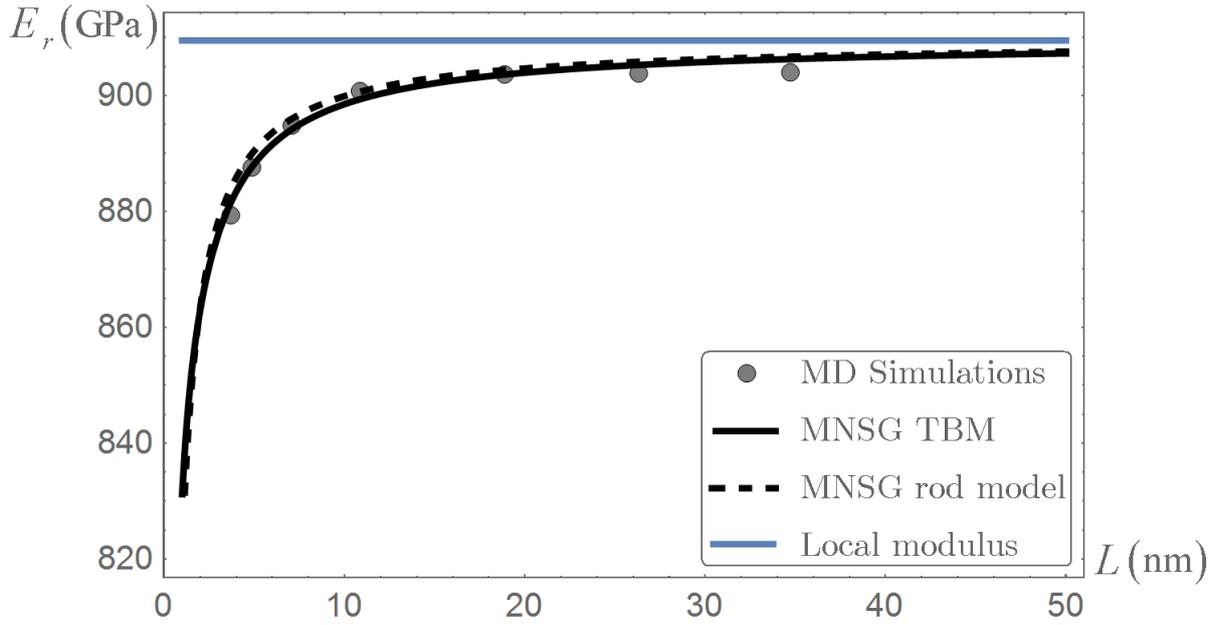

**Fig. 7.** Reduced Euler-Young elastic modulus based on the MNSG Timoshenko beam model and MNSG rod model [43] in comparison with the MD simulation results [50] of CNT $(10,10)$

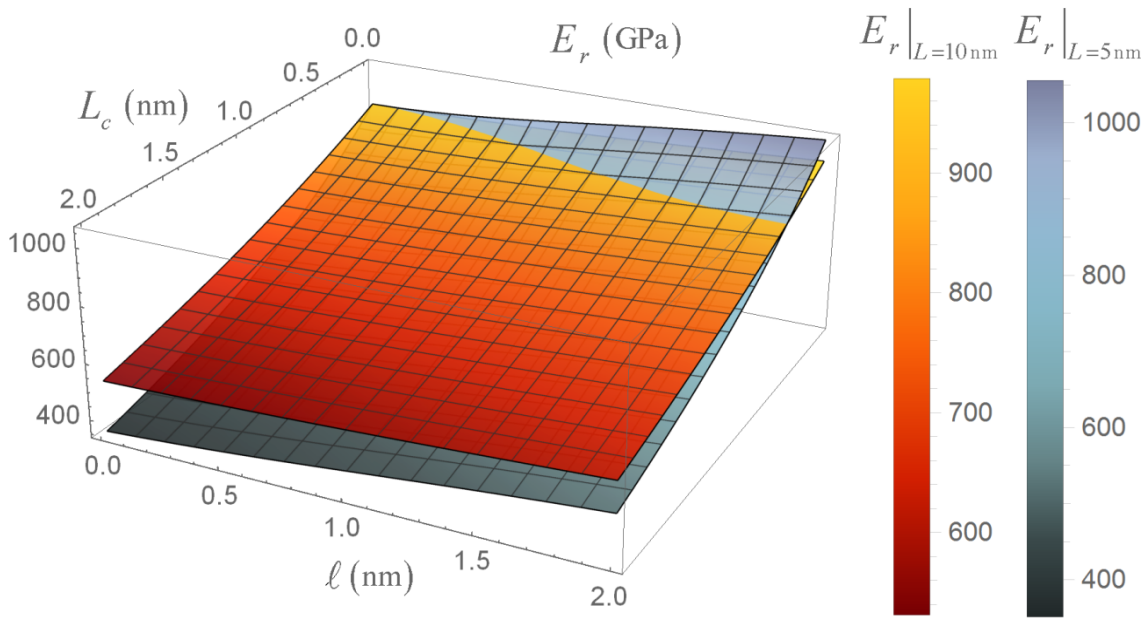

**Fig. 8.** Reduced Euler-Young elastic modulus for single-walled CNT of armchair $(10,10)$: $E_r$ vs. $L_c$ and $\ell$ for $L=5$nm and $L=10$nm

## 7. Concluding remarks

Due to the importance of carbon nanotubes having extensive applications in modern NEMS, structural responses and size-dependent constitutive behaviours of CNTs have been examined in the framework of modified nonlocal strain gradient elasticity. A novel variationally consistent approach, based on properly chosen test fields, has been exploited to introduce the nonlocal strain gradient formulation of Timoshenko elastic nano-beams. Nonlocal convolution integrals have been conveniently replaced with nonlocal differential constitutive laws supplemented with novel (non-classical) constitutive boundary conditions. An efficient analytical procedure has been illustrated and employed to investigate the elastic flexure of stubby nano-beams and closed-form nonlocal solutions for nano-cantilever and simply supported nano-beams have been provided.

The flexural response of Timoshenko nano-beam in the framework of modified nonlocal strain gradient model has been comprehensively examined and compared with the counterpart results of the classical nonlocal strain gradient (NSG) model. The drawbacks of NSG have been highlighted in case of Timoshenko nano-beams with simply supported ends. In particular, it has been shown that by prescribing inappropriately higher-order boundary conditions of strain gradient elasticity, local flexural responses are associated with identical nonlocal and gradient characteristic parameters. On the contrary, MNSG can efficiently predict softening or stiffening structural effects in terms of nonlocal and gradient parameters. The effects of transverse shear deformation on the flexural response of nano-beams have been also studied and it has been shown that shear contributions are more significant for beams with lower slenderness ratios. Similarly, it has been proven that transversal deflections of nano-beam tend to the local response as the structural length is increasing, and consequently, small-scale size-effects are more prominent for stubby nano-beams.

Based on the introduced variationally consistent nonlocal strain gradient model, the presented formulation for the reduced Euler-Young elastic modulus has been proposed to capture the small-scale behavior of elastic modulus of CNTs. It is well-established that classical continuum theory fails to predict the size-dependent constitutive properties of short CNTs. The proposed reduced elastic modulus comprises a nonlocal parameter and a material length-scale parameter to suitably exhibit the nanoscopic effects of nonlocality and nanostructure-dependent gradients. Accordingly, the novel formulation can serve as a foundation to suitably connect continuum mechanics with experimental results or molecular dynamic simulations. The reduced Euler-Young modulus has been shown to be dependent on length and diameter of CNTs, and therefore, it can successfully incorporate geometrical properties of CNTs.

The size-dependent elastic modulus of single-walled CNTs of armchair $(10,10)$ has been investigated by the MNSG Timoshenko nano-beam model. Both nonlocal and gradient characteristic parameters of the reduced Euler-Young elastic modulus have been tuned to have the best agreement with MD simulations results, while effectively incorporating the geometrical properties of CNTs. It has been demonstrated that the reduced Euler-Young modulus can exhibit both stiffness-hardening and stiffness-softening behaviours in terms of gradient and nonlocal parameters, respectively. The local elastic modulus is recovered as nonlocal and gradient parameters are vanishing. The size-dependent material response of CNTs has been also inspected in terms of CNT length. It has been shown that the reduced Euler-Young modulus provides the local modulus as the length of the single-walled CNT increases, in consequence of the dominance of small-scale effects in stubby nano-beams.

The variationally consistent nonlocal strain gradient model, which improves previous size-dependent formulations in literature, provides a viable approach for design and optimization of structural elements of advanced NEMS.

**Acknowledgements**

The financial support of the Italian Ministry for University and Research (P.R.I.N. National Grant 2017, Project code 2017J4EAYB; University of Naples Federico II Research Unit) is gratefully acknowledged.

**Table 1.** Dimensionless maximum deflections of a nano-cantilever subject to a transverse applied force at the free end point and slenderness ratio $L/h = 10$

| $L_c$ nm | Beam Model | $v^*$ (MNSG) | | | | |
|---|---|---|---|---|---|---|
| | | $\ell = 0$ nm | $\ell = 0.5$ nm | $\ell = 1$ nm | $\ell = 1.5$ nm | $\ell = 2$ nm |
| $0^+$ | BEM | 0.333333 | 0.331083 | 0.325333 | 0.317566 | 0.309119 |
| | TBM | 0.335933 | 0.333683 | 0.327933 | 0.320166 | 0.311719 |
| 0.5 | BEM | 0.385833 | 0.380833 | 0.372334 | 0.361837 | 0.350766 |
| | TBM | 0.388693 | 0.383693 | 0.375194 | 0.364697 | 0.353626 |
| 1.0 | BEM | 0.443333 | 0.435083 | 0.423334 | 0.409612 | 0.39544 |
| | TBM | 0.446452 | 0.438203 | 0.426454 | 0.412732 | 0.39856 |
| 1.5 | BEM | 0.505833 | 0.493833 | 0.478335 | 0.460891 | 0.443141 |
| | TBM | 0.509211 | 0.497213 | 0.481715 | 0.464271 | 0.446521 |
| 2.0 | BEM | 0.573333 | 0.557083 | 0.537336 | 0.515673 | 0.493869 |
| | TBM | 0.576971 | 0.560723 | 0.540976 | 0.519313 | 0.497509 |

**Table 2.** Dimensionless maximum deflections of a nano-cantilever subject to a transverse applied force at the free end point and slenderness ratio $L/h = 20$

| $L_c$ nm | Beam Model | $v^*$ (MNSG) | | | | |
|---|---|---|---|---|---|---|
| | | $\ell = 0\ nm$ | $\ell = 0.5\ nm$ | $\ell = 1\ nm$ | $\ell = 1.5\ nm$ | $\ell = 2\ nm$ |
| $0^+$ | BEM | 0.333333 | 0.33274 | 0.331083 | 0.328552 | 0.325333 |
| | TBM | 0.333983 | 0.33339 | 0.331733 | 0.329202 | 0.325983 |
| 0.5 | BEM | 0.358958 | 0.357708 | 0.355396 | 0.352208 | 0.348333 |
| | TBM | 0.359641 | 0.358391 | 0.356078 | 0.352891 | 0.349016 |
| 1.0 | BEM | 0.385833 | 0.383865 | 0.380833 | 0.376927 | 0.372334 |
| | TBM | 0.386548 | 0.38458 | 0.381548 | 0.377642 | 0.373049 |
| 1.5 | BEM | 0.413958 | 0.411208 | 0.407396 | 0.402708 | 0.397334 |
| | TBM | 0.414705 | 0.411956 | 0.408143 | 0.403456 | 0.398081 |
| 2.0 | BEM | 0.443333 | 0.43974 | 0.435083 | 0.429552 | 0.423334 |
| | TBM | 0.444113 | 0.44052 | 0.435863 | 0.430332 | 0.424114 |

**Table 3.** Dimensionless maximum deflections of a nano-cantilever subject to a transverse applied force at the free end point and slenderness ratio $L/h = 50$

| $L_c$ nm | Beam Model | $v^*$ (MNSG) | | | | |
|---|---|---|---|---|---|---|
| | | $\ell = 0$ nm | $\ell = 0.5$ nm | $\ell = 1$ nm | $\ell = 1.5$ nm | $\ell = 2$ nm |
| $0^+$ | BEM | 0.333333 | 0.333235 | 0.332949 | 0.332487 | 0.331861 |
| | TBM | 0.333437 | 0.333339 | 0.333053 | 0.332591 | 0.331965 |
| 0.5 | BEM | 0.343433 | 0.343233 | 0.342845 | 0.342281 | 0.341553 |
| | TBM | 0.343539 | 0.343339 | 0.342951 | 0.342387 | 0.341659 |
| 1.0 | BEM | 0.353733 | 0.353427 | 0.352933 | 0.352263 | 0.351429 |
| | TBM | 0.353841 | 0.353535 | 0.353041 | 0.352371 | 0.351537 |
| 1.5 | BEM | 0.364233 | 0.363817 | 0.363213 | 0.362433 | 0.361489 |
| | TBM | 0.364344 | 0.363928 | 0.363324 | 0.362544 | 0.3616 |
| 2.0 | BEM | 0.374933 | 0.374403 | 0.373685 | 0.372791 | 0.371733 |
| | TBM | 0.375046 | 0.374516 | 0.373798 | 0.372904 | 0.371846 |

**Table 4.** Dimensionless maximum deflections of a simply supported nanobeam subject to a transverse uniform load and slenderness ratio $L/h = 10$

| $L_c$ nm | Beam Model | $v^*$ (MNSG) | | | $v^*$ (NSG) | | |
|---|---|---|---|---|---|---|---|
| | | $\ell = 0\ nm$ | $\ell = 1\ nm$ | $\ell = 2\ nm$ | $\ell = 0\ nm$ | $\ell = 1\ nm$ | $\ell = 2\ nm$ |
| $0^+$ | BEM | 1.30208 | 1.22641 | 1.1414 | 1.3021 | 1.1870 | 0.9360 |
| | TBM | 1.33458 | 1.25635 | 1.16519 | 1.3346 | 1.2169 | 0.9598 |
| 1.0 | BEM | 1.42708 | 1.30208 | 1.18157 | 1.4271 | 1.3021 | 1.0275 |
| | TBM | 1.47518 | 1.34741 | 1.21842 | 1.4622 | 1.3346 | 1.0535 |
| 2.0 | BEM | 1.80208 | 1.52909 | 1.30208 | 1.8021 | 1.6475 | 1.3021 |
| | TBM | 1.87096 | 1.59494 | 1.35634 | 1.8450 | 1.6877 | 1.3346 |

**Table 5.** Dimensionless maximum deflections of a simply supported nanobeam subject to a transverse uniform load and slenderness ratio $L/h = 20$

| $L_c$ nm | Beam Model | $v^*$ (MNSG) | | | $v^*$ (NSG) | | |
|---|---|---|---|---|---|---|---|
| | | $\ell = 0\ nm$ | $\ell = 1\ nm$ | $\ell = 2\ nm$ | $\ell = 0\ nm$ | $\ell = 1\ nm$ | $\ell = 2\ nm$ |
| $0^+$ | BEM | 1.30208 | 1.27708 | 1.22641 | 1.3021 | 1.2715 | 1.1870 |
| | TBM | 1.31021 | 1.28505 | 1.2339 | 1.3102 | 1.2794 | 1.1944 |
| 1.0 | BEM | 1.33333 | 1.30208 | 1.24533 | 1.3333 | 1.3021 | 1.2157 |
| | TBM | 1.34325 | 1.31183 | 1.25458 | 1.3416 | 1.3102 | 1.2234 |
| 2.0 | BEM | 1.42708 | 1.37709 | 1.30208 | 1.4271 | 1.3940 | 1.3021 |
| | TBM | 1.43911 | 1.38895 | 1.31341 | 1.4359 | 1.4026 | 1.3102 |

**Table 6.** Dimensionless maximum deflections of a simply supported nanobeam subject to a transverse uniform load and slenderness ratio $L/h = 50$

| $L_c$ nm | Beam Model | $v^*$ (MNSG) | | | $v^*$ (NSG) | | |
|---|---|---|---|---|---|---|---|
| | | $\ell = 0$ nm | $\ell = 1$ nm | $\ell = 2$ nm | $\ell = 0$ nm | $\ell = 1$ nm | $\ell = 2$ nm |
| $0^+$ | BEM | 1.30208 | 1.29748 | 1.28528 | 1.3021 | 1.2971 | 1.2823 |
| | TBM | 1.30338 | 1.29878 | 1.28657 | 1.3034 | 1.2984 | 1.2836 |
| 1.0 | BEM | 1.30708 | 1.30208 | 1.28948 | 1.3071 | 1.3021 | 1.2873 |
| | TBM | 1.30849 | 1.30349 | 1.29087 | 1.3084 | 1.3034 | 1.2886 |
| 2.0 | BEM | 1.32208 | 1.31588 | 1.30208 | 1.3221 | 1.3170 | 1.3021 |
| | TBM | 1.32361 | 1.3174 | 1.30359 | 1.3234 | 1.3184 | 1.3034 |

**Table 7.** Reduced Euler-Young elastic modulus for single-walled CNT of armchair $(10,10)$

| $L_c$ nm | $E_r\|_{L=5\text{nm}}$ | | | | | $E_r\|_{L=10\text{nm}}$ | | | | |
|---|---|---|---|---|---|---|---|---|---|---|
| | $\ell=0$ nm | $\ell=0.5$ nm | $\ell=1$ nm | $\ell=1.5$ nm | $\ell=2$ nm | $\ell=0$ nm | $\ell=0.5$ nm | $\ell=1$ nm | $\ell=1.5$ nm | $\ell=2$ nm |
| $0^+$ | 909.5 | 929.594 | 973.169 | 1018.84 | 1056.25 | 909.5 | 915.514 | 931.251 | 953.386 | 978.686 |
| 0.5 | 690.542 | 720.084 | 765.778 | 812.638 | 851.483 | 786.794 | 796.857 | 814.567 | 837.554 | 863.248 |
| 1.0 | 538.739 | 571.674 | 617.239 | 664.083 | 703.637 | 685.634 | 698.313 | 717.201 | 740.598 | 766.419 |
| 1.5 | 430.357 | 463.542 | 507.526 | 553.301 | 592.751 | 601.656 | 615.93 | 635.4 | 658.842 | 684.538 |
| 2.0 | 350.812 | 382.718 | 424.352 | 468.372 | 507.099 | 531.433 | 546.585 | 566.202 | 589.408 | 614.77 |